\documentclass[aps,twocolumn, showpacs,preprintnumbers,amsmath,amssymb,prd,superscriptaddress]{revtex4-2}
\usepackage[utf8]{inputenc} 
\usepackage{CJK}
\usepackage{amsmath}
\usepackage{amssymb}
\usepackage{bm}
\usepackage{graphicx}
\usepackage[justification=raggedright,singlelinecheck=false]{caption}
\usepackage{subcaption}
\usepackage{mathrsfs} 
\usepackage[mathscr]{euscript}
\usepackage{multirow}
\usepackage[normalem]{ulem}
\usepackage{graphicx}
\usepackage{booktabs}
\usepackage{xcolor, color, framed}
\usepackage[colorlinks, linkcolor=red,anchorcolor=green,citecolor=blue]{hyperref}
\usepackage{slashed}
\usepackage{ulem}

\graphicspath{{./figures/}}
\usepackage{tikz,xcolor,hyperref}
\usepackage{ctable}
\usepackage{pgffor}  
\usepackage{avant}   

\definecolor{lime}{HTML}{A6CE39}
\DeclareRobustCommand{\orcidicon}{
	\begin{tikzpicture}
	\draw[lime, fill=lime] (0,0) 
	circle [radius=0.16] 
	node[white] {{\fontfamily{qag}\selectfont \tiny ID}};
	\draw[white, fill=white] (-0.0625,0.095) 
	circle [radius=0.007];
	\end{tikzpicture}
	\hspace{-2mm}
}

\foreach \x in {A, ..., Z}{%
	\expandafter\xdef\csname orcid\x\endcsname{\noexpand\href{https://orcid.org/\csname orcidauthor\x\endcsname}{\noexpand\orcidicon}}
}

\renewcommand{\sout}{\bgroup \color{red} \ULdepth=-.5ex \ULset}

\begin{document}
\begin{CJK*} {UTF8} {gbsn}
\begin{abstract}
Measurement of particle emission source is a fundamental objective of femtoscopy in high-energy nuclear collisions. 
Conventional analyses rely on Gaussian {parameterizations} of pair emission sources, which makes the extraction of single-particle emission sources challenging, particularly for rare particles.
Here, we introduce a novel 
{statistical reconstruction method} 
that allows extracting information of the target source relative to a data-constrained reference source instead of the Gaussian assumption.
The correlation function is expressed as an ensemble average over {the} 
single-particle-conditioned correlation kernel,
defined as the particle-by-particle contribution to the correlation function conditioned by the target particles. 
For particles with rare {yields}, the particle-by-particle distribution of this kernel can be transformed into event-by-event extraction and becomes experimentally accessible, enabling a direct statistical reconstruction of the emission source of single particles, instead of inferring a pair source. 
We apply this method to reconstruct $J/\psi$ source via {$p$-$J/\psi$} correlations, using HAL QCD-derived $NJ/\psi$ potentials in $\sqrt{s}=13.6$~TeV $pp$ collisions simulated with EPOS4HQ. The reconstructed source reproduces the key characteristics and this new {approach} achieves a systematic uncertainty of approximately $13\%$ based on EPOS4 simulation.
\end{abstract}

\title{Reconstructing sources of rare particles by femtoscopic correlations}
\author{Liang Zhang(张良)\orcidA{}}
\affiliation{Key Laboratory of Nuclear Physics and Ion-beam Application (MOE), Institute of Modern Physics, Fudan University, Shanghai 200433, China}
\affiliation{Shanghai Institute of Applied Physics (SINAP), Shanghai, China}
\author{Song Zhang(张松)\orcidB{}}
\email{song\_zhang@fudan.edu.cn}
\affiliation{Key Laboratory of Nuclear Physics and Ion-beam Application (MOE), Institute of Modern Physics, Fudan University, Shanghai 200433, China}
\affiliation{Shanghai Research Center for Theoretical Nuclear Physics, NSFC and Fudan University, Shanghai 200438, China}
\author{Kai-Jia Sun(孙开佳)\orcidD{}}
\email{kjsun@fudan.edu.cn}
\affiliation{Key Laboratory of Nuclear Physics and Ion-beam Application (MOE), Institute of Modern Physics, Fudan University, Shanghai 200433, China}
\affiliation{Shanghai Research Center for Theoretical Nuclear Physics, NSFC and Fudan University, Shanghai 200438, China}
\author{Yu-Gang Ma(马余刚)\orcidC{}}
\email{mayugang@fudan.edu.cn}
\affiliation{Key Laboratory of Nuclear Physics and Ion-beam Application (MOE), Institute of Modern Physics, Fudan University, Shanghai 200433, China}
\affiliation{Shanghai Research Center for Theoretical Nuclear Physics, NSFC and Fudan University, Shanghai 200438, China}
\affiliation{School of Physics, East China Normal University, Shanghai 200241, China}
\maketitle

\section{Introduction}
\label{sec:intro}

Hanbury Brown and Twiss (HBT) {first} pioneered identical-particle intensity interferometry to measure the apparent angular diameter of remote stars~\cite{HanburyBrown:1956bqd}. 
In the 1960s, this concept was extended to high-energy physics~\cite{Goldhaber:1960sf},
where HBT interferometry has since developed into a powerful tool for exploring the femtometer-scale spatiotemporal structure of particle-emission sources \cite{Kopylov:1974th,Zajc:1984vb} in intermediate-energy  \cite{Boal,Bauer} and high-energy \cite{Lisa:2005dd,Heinz:1999rw,Wiedemann:1999qn} heavy-ion collisions, {which} provides an ideal venue to study {properties} of nuclear  matter~ \cite{Chen:NST2024,Shou:NST2024,Lu:NST2025,Wang:NST2025,Yu:NST2025,Dong:NST2025}.

At such microscopic scales, correlation analyses must account for final-state interactions. This inclusion allows the method to be extended to non-identical particle pairs~\cite{lednicky_final_1981,Pratt:1986cc,Bowler:1991vx,lednicky_influence_1996,lednicky_femtoscopy_2001,Lednicky:2002fq,WeiYB-2004,MaYG-2012,MaYG-2015}. In the context of relativistic heavy-ion collisions, this methodology is commonly referred to as ``femtoscopy'' \cite{lednicky_femtoscopy_2001,Lednicky:2002fq,Lisa:2005dd,Heinz:1999rw,Wiedemann:1999qn}. It is widely employed to characterize the geometry of emission sources~\cite{Koonin:1977fh,Zajc:1984vb,Pratt:1986cc,He:2020jzd,ALICE:2020ibs,ALICE:2023sjd,Xu:2024dnd,Xiong:2025bmd,Wang:2024bpl,Xi:2025iwd}
and dynamics of system evolution~\cite{Wang:2021mrv,Wang:2023ygv,Qiao:2024wha,Wang:2024yke,Wang:2025htd,Ma:2006zq}
and, more recently, to probe the nature of hadronic interactions at low relative momenta~\cite{lednicky_final_1981,STAR:2015kha,Haidenbauer:2018jvl,Fabbietti:2020bfg,ALICE:2022uso,ALICE:2021cpv,Si:2025eou}, including studies of exotic hadrons and hadronic resonant states~\cite{STAR-LL-2015,Kamiya:2019uiw,ALICE:2023wjz,ALICE:2025aur,ALICE:2025byl,SunKJ:NST2026,Liu:2024uxn,Shou:NST2024,Chen:NST2025}. 

The experimental measurements {from} the ALICE Collaboration {have} demonstrated that the emission sources of $\pi$, $K$, $p$ and $\Lambda$ in $13$~TeV $pp$ collisions at the {CERN Large Hadron Collider (LHC)} originate from a common hadron-emission source~\cite{ALICE:2020ibs,ALICE:2023sjd}. 
These findings naturally raise the question of whether the emission sources of short-lived charm hadrons exhibit a similar spatial structure. Since charm hadrons are mainly produced at the early stage of the collision, their emission sources are expected to be more compact than those of longer-lived particles mentioned above. 
More fundamentally, conventional femtoscopy accesses source information only through two-particle observables, such that the single-particle emission source of charm hadrons is not directly measurable. 
As these hadrons are emitted during a non-equilibrium state, the commonly adopted Gaussian {parametrizations} should not be assumed a priori. Consequently, the extraction of charm-hadron emission sources within the standard femtoscopy framework remains an open problem. 

With the continuous improvement of experimental precision, particle source extraction in femtoscopy has moved beyond traditional Gaussian parametrizations~\cite{Lisa:2005dd,Koonin:1977fh,Zajc:1984vb}. Since the particle emission source is inferred from measured two-particle correlation functions, source reconstruction constitutes an ill-posed inverse problem, motivating the development of various inverse approaches, including machine learning techniques~\cite{Wang:2024bpl}, optical deblurring algorithms~\cite{Xu:2024dnd}, and regularization methods such as Tikhonov regularization~\cite{Xiong:2025bmd}. While these methods significantly improve the reconstruction of femtoscopic sources, they are all formulated at the level of two-particle (pair) sources that are directly accessible through correlation measurements. In contrast, the direct extraction of a source of single particles, especially for rare particles, remains an open problem, as the single-particle emission information is implicitly folded into pair observables. 

In the present study, we introduce a {statistical reconstruction method}, 
which reformulates femtoscopy as an inverse problem for emission sources of single {particles}.
The method is based on expressing the measured two-particle correlation as an ensemble average over single-particle-conditioned correlation kernels, whose distribution encodes the underlying source of target particles. 
This reconstruction cannot be achieved within conventional source imaging approaches, which invert only the source of pairs. 
In practice, the reconstruction requires (i) a well-characterized emission source of an abundant reference hadron (e.g., pions or protons) which can be obtained by phenomenological models, and (ii) a constrained interaction potential, provided by effective field theories or lattice QCD.

As a proof of concept, we demonstrate the method through EPOS4HQ simulations~\cite{Werner:2023fne,Werner:2023zvo,Werner:2023jps,Werner:2023mod,Zhao:2023ucp,Zhao:2024ecc}, where the $J/\psi$ emission source is reconstructed via $p$-$J/\psi$ correlations with $NJ/\psi$ potentials derived from HAL QCD calculations~\cite{Lyu:2024ttm}. In this work, we focus on validating the stability and statistical uncertainty of the reconstruction method under controlled input conditions.

This paper is structured as follows. {Section} \ref{sec:method} introduces our {statistical reconstruction method}.  
{Section} \ref{sec:reconstruct} shows the reconstruction of the $J/\psi$ source in EPOS4HQ {simulations} to validate the method. Finally, a summary is provided in {Section} \ref{sec:summary}.

\section{Statistical reconstruction method}
\label{sec:method}
\subsection{Methodology}
\label{subsec:method}

The two-particle momentum correlation function constitutes the fundamental observable in femtoscopy. It is defined as~\cite{Lisa:2005dd}
\begin{equation}
\label{eq:correlation}
    C(\mathbf{k}^*)=\int d^3\mathbf{r}_1 d^3 \mathbf{r}_2 S_1(\mathbf{r}_1)S_2(\mathbf{r}_2) \left|\psi_{\mathbf{k}^*}(\mathbf{r}_1-\mathbf{r}_2)\right|^2,
\end{equation}
where $\mathbf{k}^*=\frac{1}{2}(\mathbf{p}_1-\mathbf{p}_2)$ denotes the relative momentum in the pair rest frame, and $\mathbf{r}_i$ {are} the particle coordinates. $S_1(\mathbf{r}_1)$ and $S_2(\mathbf{{r}_2})$ {denote} the normalized emission sources of single particles, while $\left|\psi_{\mathbf{k}^*}(\mathbf{r}_1-\mathbf{r}_2)\right|^2$ incorporates contributions from {quantum statistics} (for identical {particles}) and {final-state interactions}. 
This formalism provides the basis for extracting source properties from experimental data, allowing a quantitative investigation of the space-time structure of particle-emitting sources. 

In conventional approaches, particle emission sources are typically modeled as Gaussian distributions~\cite{Koonin:1977fh,Zajc:1984vb,ALICE:2022uso,ALICE:2021cpv,Fabbietti:2020bfg}. Recently, the ALICE collaboration proposed the Resonance Source Model (RSM)~\cite{ALICE:2020ibs,ALICE:2023sjd}, which incorporates a universal Gaussian core for primordially produced particles, along with an exponential tail from resonance-decay contributions. The two-particle Gaussian source can be expressed as:
\begin{equation}
\label{eq:two-particle_Gaussian_source}
    S_{12}(r)=\frac{1}{(4\pi R^2)^{3/2}}\exp{\left(-\frac{r^2}{4R^2}\right)}.
\end{equation}
This Gaussian parametrization continues to be widely adopted in femtoscopic analyses~\cite{STAR:2015kha,ALICE:2020mfd,ALICE:2019gcn,ALICE:2021cpv,ALICE:2023sjd}. 
Motivated by these developments, it is necessary to test the universality of the Gaussian source for early-stage produced particles. In particular, one can investigate the $J/\psi$ emission source via $p$-$J/\psi$ correlations without assuming a Gaussian form.

Building on the conventional approach, pair source distributions are often inferred by fitting correlation functions with predetermined interaction wavefunctions. To isolate the source distribution of an individual particle, $S_1(\mathbf{r}_1)$, in {Eq.} \eqref{eq:correlation}, the integral can be rewritten as
\begin{equation}
    \label{eq:single-source-integral}
    C(k^*)=\int dr_1 {S_{1}}(\mathbf{r}_1) \tilde{C}_{k^*}(\mathbf{r}_1),
\end{equation}
where $\tilde{C}_{k^*}(\mathbf{r}_1)$, termed the single-particle-conditioned correlation kernel,
represents the correlation between particle 1 located at $r_1$ and the ensemble of particle 2:
\begin{equation}
\label{eq:single-particle-correlation}
    \tilde{C}_{k^*}(\mathbf{r}_1)=\int d^3\mathbf{r}_2 {S_{2}} (\mathbf{r}_2)\left|\psi_{k^*}(\mathbf{r}_1-\mathbf{r}_2)\right|^2,
\end{equation}
under the assumption of isotropic emission. For simplicity, the angular distribution of the source is taken to be uniform, and hereafter the angular-averaged source is denoted simply as $S_i(r_i)$.

{Equation} \eqref{eq:single-source-integral} can be interpreted statistically, which corresponds to the mean value $\langle\tilde{C}_{k^*}(r_1)\rangle$ over a set of samples drawn from $r_1\sim S_1(r_1)$. The probability of $y=\tilde{C}_{k^*}(r_1)$ can then be expressed as 
\begin{equation}
\label{eq:statistical-explain}
    p_{\tilde{C}_{k^*}}(y)=\int S_1 (r) \delta\left[y-\tilde{C}_{k^*}(r)\right]dr,
\end{equation}
which describes the transformation of the source distribution $S_{1}(r)$  into the distribution of $y$ through the functional map $\tilde{C}_{k^*}(r)$. This property is essential, as it enables a direct statistical reconstruction of the emission source of single particles, rather than an indirect inference of an effective source of pair. It should be noted that the extracted $S_1$ is not a distribution in absolute spacetime, but a relative distribution obtained after averaging the spatial distribution of particle 2. However, it still contains spatial information about particle 1.

Reversal of the thought of {Eq.}~\eqref{eq:statistical-explain}, a reconstruction strategy for source is presented as,
\begin{equation}
\label{eq:reconstruction-sampling}
\begin{aligned}
    S^{(k^*)}_1(r)&=\int p(y)\delta\left[r-\tilde{C}^{-1}_{k^*}(y)\right]dy\\
                                &=\sum_j {\left[p(y_j)\left|\tilde{C}'_{k^*}(r)\right|\right]_{y_j=\tilde{C}_{k^*}(r)}}.
\end{aligned}
\end{equation}
Due to the potentially non-monotonicity of $\tilde{C}_{k^*}(r)$ function, the reconstructed source distribution $S^{(k^*)}_1(r)$ at individual $k^*$ may deviate from original source distribution $S_1(r)$.
In the following, we show that a coherent combination of reconstructions over multiple relative-momentum bins provides a robust estimate of the underlying source.

This approach fundamentally relies on two essential requirements, the distribution of $y=\tilde{C}_{k^*}(r_1)$ and the analytical calculation of $\tilde{C}_{k^*}(r)$. The latter one is based on the established inputs, a known source of single particles as reference source and a well-constrained interaction potential, as introduced in {Sec.} \ref{sec:intro}. 

In experiments, the correlation function $C(k^*)$ is determined through the normalized ratio \cite{Lisa:2005dd,ALICE:2020ibs,ALICE:2023sjd}
\begin{equation}
    C(k^*)=\mathcal{N}\frac{N^\text{same}(k^*)}{N^\text{mixed}(k^*)},
\end{equation}
where $\mathcal{N}$ is the normalization factor,  $N^{\text{mixed}}(k^*)$ represents the distribution of uncorrelated pairs obtained from event mixing, and $N^\text{same}(k^*)=\sum_i^{N_\text{events}} \sum_j^{N^i_T} n^\text{same}_{ij}(k^*)$ is the distribution of correlated pairs from the same event. Here $N_\text{events}$ denotes the total number of events, $N^i_T$ is the number of target particle to be reconstructed in $i$th event and $n^\text{same}_{ij}(k^*)$ represents the pair number with momentum $k^*$ including $j$th target particle in $i$th event. The corresponding conditioned kernel
$\tilde{C}_{k^*}$ for a target particle in the $i$th event can then be defined as
\begin{equation}
\label{eq:measure_Ctilde}
    \tilde{C}_{k^*;ij}=\mathcal{N}N_T^{\rm{tot}}\frac{n_{ij}^\text{same}(k^*)}{N^\text{mixed}(k^*)}.
\end{equation}
This {definition} allows the particle-by-particle distribution of $y=\tilde{C}_{k^*}$ to be directly accessed experimentally, providing the statistical input required for reconstructing the emission source of single particles. However, experimentally determining the conditioned kernel
is challenging because it is difficult to isolate individual particles for pairing in a given event. Equation~\ref{eq:measure_Ctilde} also incorporates particle production fluctuations into the kernel fluctuations, which {increase} the {systemaic uncertainty}.

\begin{figure}
    \centering
    \includegraphics[width=0.9\linewidth]{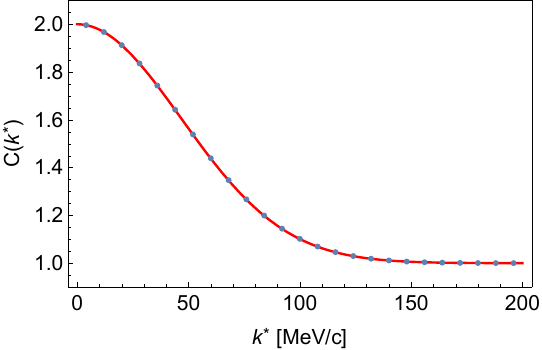}
    \caption{The analytically computed correlation function derived from Gaussian emission sources with radii $R_1$ = 0.5~fm and $R_2 = 1.4$~fm for $S_1$ and $S_2$ correspondingly, employs the wavefunction $\psi_{\mathbf{k}^*}(\mathbf{r})$ = $\sqrt{2}\cos(\mathbf{k}^* \cdot \mathbf{r})$. Blue markers denote candidate momenta selected for reconstruction validation.}
    \label{fig:Validation_correlation}
\end{figure}

\subsection{Validation}
\label{subseq:simple_validation}

\begin{figure*}
    \centering
    \includegraphics[width=1\textwidth]{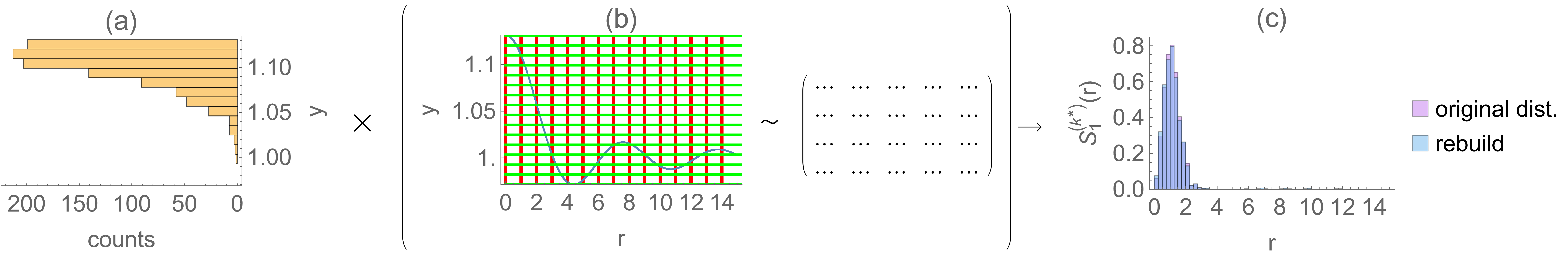}
    \caption{Flowchart on reconstructing $S^{(k^*)}_1(r)$. {Panel} (a) shows the histogram of $y=\tilde{C}_{k^*}(r)$ where $k^*=100{~\mathrm{MeV}/c}$. {Panel} (b) is the $\tilde{C}_{k^*}(r)$ function and presents the calculation of {the} reconstruction kernel $K_{k^*}$. {Panel} (c) is the reconstruction of $S^{(k^*)}_1(r)$ and {comparison} with the ground-truth source distribution.}
    \label{fig:Validation_result}
\end{figure*}

To validate the feasibility of our {statistical reconstruction method}, 
we perform a controlled numerical test. We use predefined Gaussian emission sources with radii $R_1=0.5$~fm and $R_2=1.4$~fm for $S_1$ and $S_2$ correspondingly, and adopt the wave function
\begin{equation}
\psi_{\mathbf{k}^*}(\mathbf{r})=\sqrt{2}\cos(\mathbf{k}^*\cdot \mathbf{r}).    
\end{equation}
The corresponding analytical correlation function is shown in {Fig.} \ref{fig:Validation_correlation}.  We then quantitatively evaluate the reconstruction accuracy by comparing the extracted source profiles against ground-truth inputs. 

The validation {is initiated} by {sampling} radial coordinates $r_1$ for particle 1 from {the} source distribution $S_1$. For each sampled coordinate, the $\tilde{C}_{k^*}(r)$, defined in {Eq.} \eqref{eq:single-particle-correlation}, is evaluated over the range $r\in[0,15]$~fm. The resulting distribution of  $y=\tilde{C}_{k^*}(r)$ at $k^*=100{~\mathrm{MeV}/c}$ is shown as a histogram in {Fig.} \ref{fig:Validation_result}a. 

The {analytical} form of $\tilde{C}_{k^*}(r)$, shown as the blue curve in {Fig.} \ref{fig:Validation_result}b, is discretized on a {two-
dimensional} grid with 200 bins in $y$ and 60 bins in $r$. Within each bin, the local slope $d\tilde{C}_{k^*}$ $(r)/dr$ is calculated. These slopes are used to construct the reconstruction kernel
$K_{k^*}$, where the elements $(K_{k^*})_{ij}$ are set to zero if the curve of $\tilde{C}_{k^*}(r)$ does not pass through the bin $(r_i,y_j)$.

\begin{figure*}
    \centering
    \includegraphics[width=0.8\linewidth]{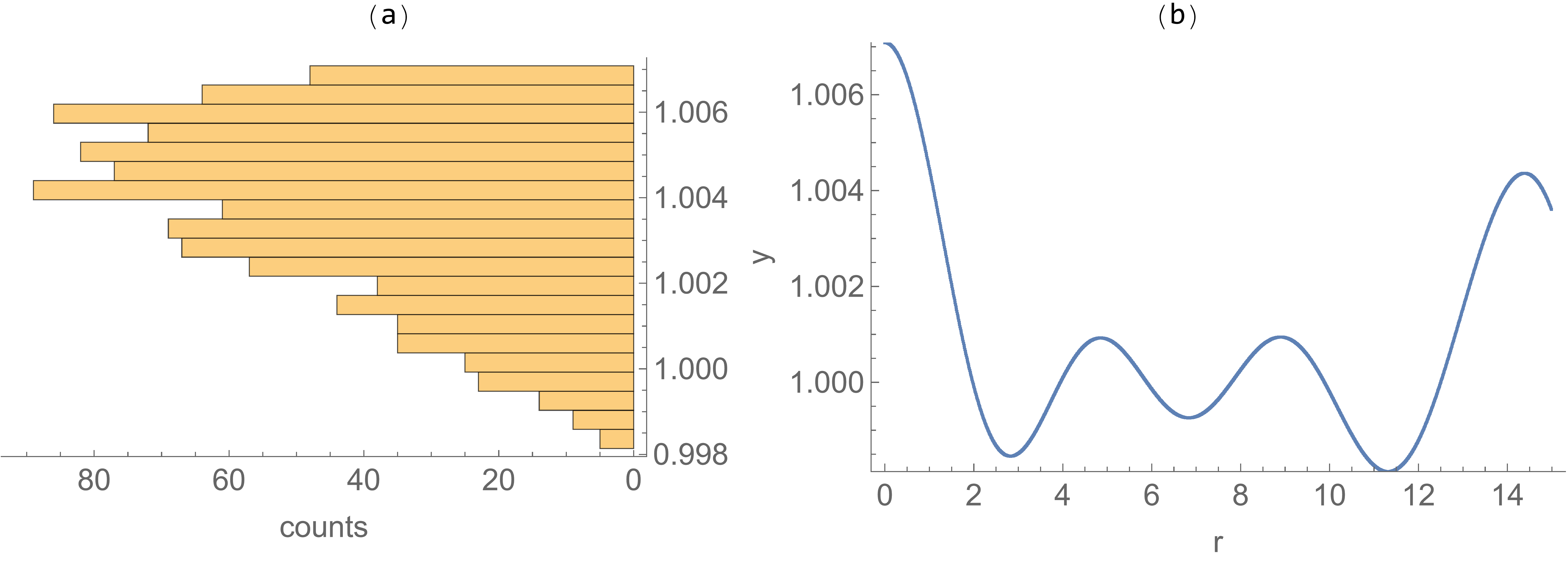}
    \caption{{Panel} (a) presents the histogram of $y=\tilde{C}_{k^*}(r)$ where $k^*=156{~\mathrm{MeV}/c}$. {Panel} (b) illustrates the nonmonotonic behavior of $\tilde{C}_{k^*}(r)$. This violation of bijectivity induces ambiguity in the reconstruction mapping, resulting in a {poorly} reconstructed source distribution.}
    \label{fig:Validation_nonmonotonicity}
\end{figure*}
\begin{figure}
    \centering
    \includegraphics[width=0.9\linewidth]{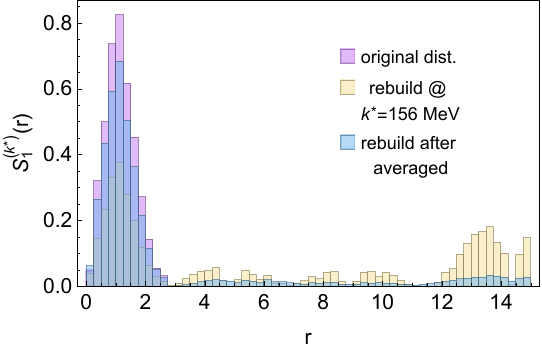}
    \caption{A bad reconstruction example (yellow) at $k^*=156{~\mathrm{MeV}/c}$ due to the nonmonotonic behavior of $\tilde{C}_{k^*}(r)$ shown in {Fig.} \ref{fig:Validation_nonmonotonicity}. Averaging on multi-momentum reconstructions (blue) can reduce the effects of these fake peaks.}
    \label{fig:Validation_bad_aveage}
\end{figure}

To address regions where $\tilde{C}_{k^*}(r)$ varies rapidly and passes through multiple $y$-bins within a single $r$-bin, we implement a multiplicity-averaging procedure:
\begin{equation}
    K_{ij}\rightarrow (K_{k^*})_{ij}/n^\text{cross $y$-bins}_i,
\end{equation}
where $n^\text{cross $y$-bins}_i$ counts the number of $y$-bins traversed by the curve within the $i$th $r$-bin. This prevents overcounting of probability when the reconstruction mapping is multivalued. 

The reconstructed source distribution at a given $k^*$ is then obtained via
\begin{equation}
    S^{(k^*)}_1(r_i)=\frac{ K_{k^*}\cdot \boldsymbol{y}^\text{counts}}{\sum_j y^\text{counts}_j}\Delta r_i,
\end{equation}
where $\boldsymbol{y}^\text{counts}$ denotes the histogram-count vector of $\tilde{C}_{k^*}(r)$. As demonstrated in {Fig.~\ref{fig:Validation_result}-(c)}, the reconstructed distribution agrees excellently with the ground-truth source.

However, not every reconstruction yields a success story. The nonmonotonic behavior of $\tilde{C}_{k^*}(r)$ ({Fig.} \ref{fig:Validation_nonmonotonicity}) violates the bijectivity of reconstruction mapping. As a result, the reconstructed source distributions exhibit pronounced artifacts. For example, at $k^*=156{~\mathrm{MeV}/c}$ ({Fig.} \ref{fig:Validation_bad_aveage}), spurious peaks arise in reconstruction because $\tilde{C}_{k^*}(r)$ {possesses} non-zero slopes within $r<3$~fm {[Fig. \ref{fig:Validation_nonmonotonicity}-(b)]}. To suppress these artifacts, we average reconstructions performed at multiple momenta. This multi-momentum averaging amplifies the contribution from the true source and yields a significantly improved profile, as shown in {Fig.} \ref{fig:Validation_bad_aveage}.

\section{Reconstruct \texorpdfstring{$J/\psi$}{Jpsi} source from EPOS4HQ}
\label{sec:reconstruct}

\subsection{Model setup}
EPOS4 is a comprehensive simulation framework for high-energy particle collisions~\cite{Werner:2023zvo,Werner:2023fne,Werner:2023jps,Werner:2023mod}. It provides a fully self-consistent description of both proton-proton ($pp$) and nucleus-nucleus ({$AA$}) collisions by integrating four core concepts: parallel scattering (multiple partons {or} nucleons acting parallelly in high-energy collisions), energy conservation, factorization (separating hard processes into parton distribution functions and differential scattering cross sections), and {the} dynamic saturation effect (screening of {low-transverse-momentum} particle production by saturation scale). The framework first models the initial particle interactions using an {$S$}-matrix approach to handle parallel scatterings. Subsequently, the collision evolution is treated through a sequence of secondary processes, beginning with core‑corona separation and followed by hydrodynamic evolution to describe collective behavior, microcanonical hadronization to convert the fluid into hadrons, and finally the UrQMD hadron cascade acting as an afterburner to simulate resonant decays and hadronic rescatterings. 

In pure EPOS4 framework, heavy flavor partons are created in primary interactions. Their subsequent interactions with the simulated quark-gluon plasma are incorporated via the dedicated HQ module. This module also enables, under specific conditions, the coalescence of plasma partons with heavy-flavor partons to form heavy-flavor hadrons. The extended version, EPOS4HQ \cite{Zhao:2023ucp,Zhao:2024ecc}, is endowed with the capability to describe a wide range of heavy-flavor observables from experiments and remains accessible for further studies.

To compute the correlation function in {Eq.} \eqref{eq:correlation}, we {simulate} $10^5$ $pp$ collisions at $\sqrt{s}=13.6$~TeV within EPOS4HQ, which {provide} the spatiotemporal emission sources for protons and $J/\psi$. Eventwise particle samples are subjected to standard fiducial selections; protons are required to satisfy a pseudorapidity window of $-0.9<\eta<0.9$ and a transverse-momentum range of $0.25<p_{\mathrm{T}}<3~\mathrm{GeV}/c$, while $J/\psi$ candidates are selected with $-0.9<\eta<0.9$ and $p_{\mathrm{T}}>1~\mathrm{GeV}/c$. The $p$-$J/\psi$ scattering wavefunction is derived from the $N$-$J/\psi$ potential obtained via HAL QCD calculations \cite{Lyu:2024ttm}, a non-perturbative approach extracting spatial potentials from lattice QCD \cite{Ishii:2006ec, Ishii:2012ssm, Aoki:2020bew}. Final-state correlations are computed using the {correlation analysis tool using the Schr\"odinger equation}~\cite{Mihaylov:2018rva}, employing a test‑particle analysis with event mixing over all simulated collisions.

\subsection{Priori proton source}
\label{subsec:proton_source}
Since femtoscopic sources are local in momentum space, even high-momentum protons can still contribute to the $p$-$p$ correlation when their momentum directions are strongly aligned due to collective flow. While the flow effect on $J/\psi$ is weaker, especially in $pp$ collisions, the proton- and $J/\psi$-emission {processes are} {regarded} as uncorrelated and particle momentum magnitude is the primary criterion for particle selection in the $p$-$J/\psi$ correlation analysis. However, the proton source is different in $p-p$ and $p$-$J/\psi$ correlations, and we {cannot} extract proton source via $p$-$p$ correlation for $p$-$J/\psi$ pair analysis. In this simulation, the relative momentum is restricted to $0<k^*<400~\mathrm{MeV}/c$. For a pair with momenta $p_1$ for proton and $p_2$ for $J/\psi$, the constraint on $k^*$ imposes a strong kinematic constraint on the opening angle $\theta_{12}$, since $k^*=\frac{1}{2}\sqrt{p_1^2+p_2^2-2p_1p_2\cos\theta_{12}}$. When particles have large momenta, the {acceptance} of $p_1$ and $p_2$,
\begin{equation}
    \mathrm{Acc}(p_1,p_2)=\int_0^{400}\sin(\theta_{12})\frac{\partial\theta_{12}}{\partial{k}}dk/2,
\end{equation}
can be small due to a narrow $\theta_{12}$ interval. 
Therefore, pairs with large $p_1$ and $p_2$ contribute weakly to the selected $k^*$ region. This provides the practical basis for selecting protons within a controlled momentum window to construct the priori reference source in the present reconstruction framework.

\begin{figure}
    \centering
    \includegraphics[width=0.8\linewidth]{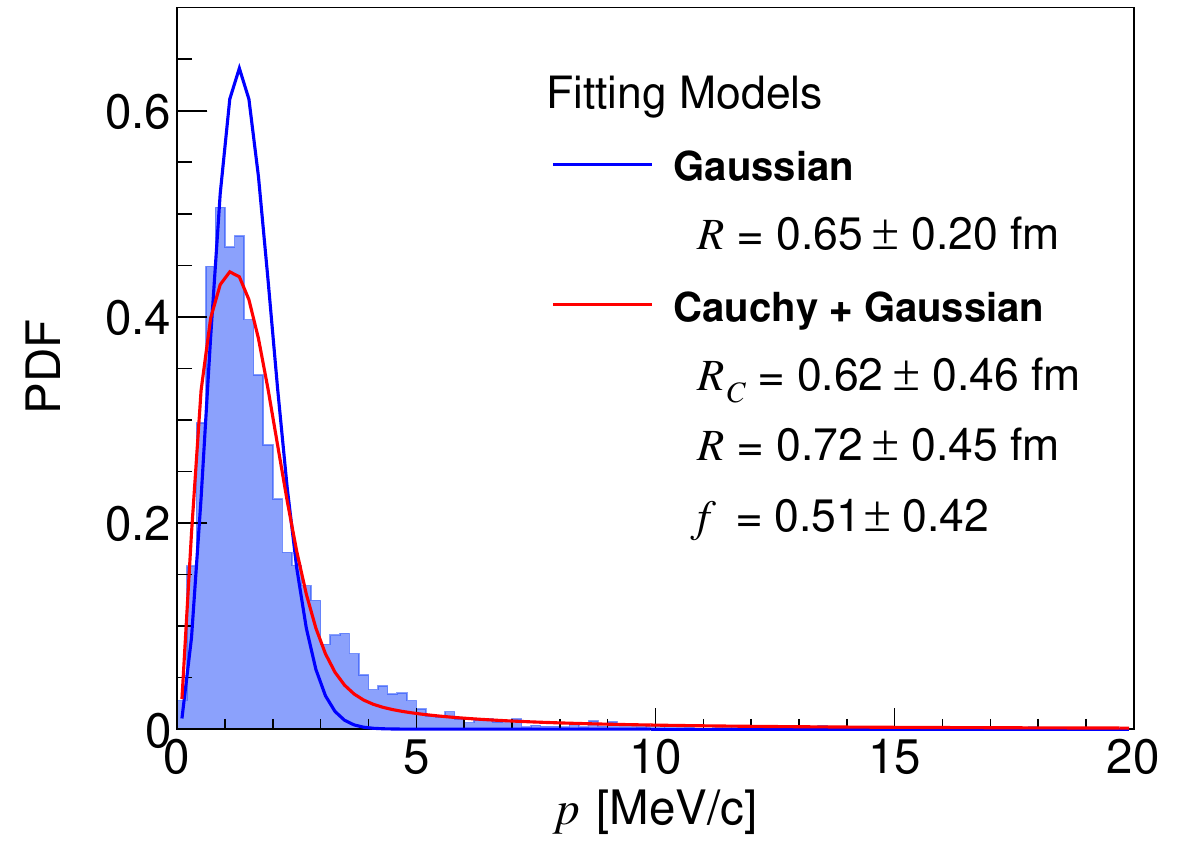}
    \caption{The protons with momentum $<400~\mathrm{MeV}/c$ are selected as the known reference source. The spatial distribution is generated by EPOS4HQ and fitted with Gaussian distribution as well as {Eq.}~\eqref{eq:CauchyGaussian_source}, where {the} Gaussian + Cauchy distribution performs better.}
    \label{fig:fit-proton-source}
    
\end{figure}

In this simulation, protons with momentum $<400~\mathrm{MeV}/c$ are selected as the known reference source. To obtain its spatial distribution, we adopt the {Gaussian} core + Cauchy tail~\cite{ALICE:2020ibs,ALICE:2023sjd}, 
\begin{equation}
    \label{eq:CauchyGaussian_source}
    S_{p}(r)=f\frac{R_{C}}{\pi^2\left(R_{C}^{2}+r^{2}\right)^{2}}+(1-f)\frac{1}{\sqrt{4\pi R^{2}}^{3}}\exp\left(-\frac{r^2}{4R^2}\right),
\end{equation}
to describe the proton source generated by EPOS4, with $R_C\simeq0.68~\mathrm{fm}$, $R\simeq0.80~\mathrm{fm}$ and the mixing fraction $f\simeq0.51$ ({Fig.}~\ref{fig:fit-proton-source}). Although this fitting introduces model dependence into our reconstruction framework, proton sources have been extensively studied in different phenomenological models and are constrained by available experimental data. The model-to-model differences are relatively small and can be incorporated as a systematic uncertainty in the present reconstruction.

\subsection{Extracting \texorpdfstring{$J/\psi$}{Jpsi} source}
\begin{figure}
    \centering
    \includegraphics[width=1\linewidth]{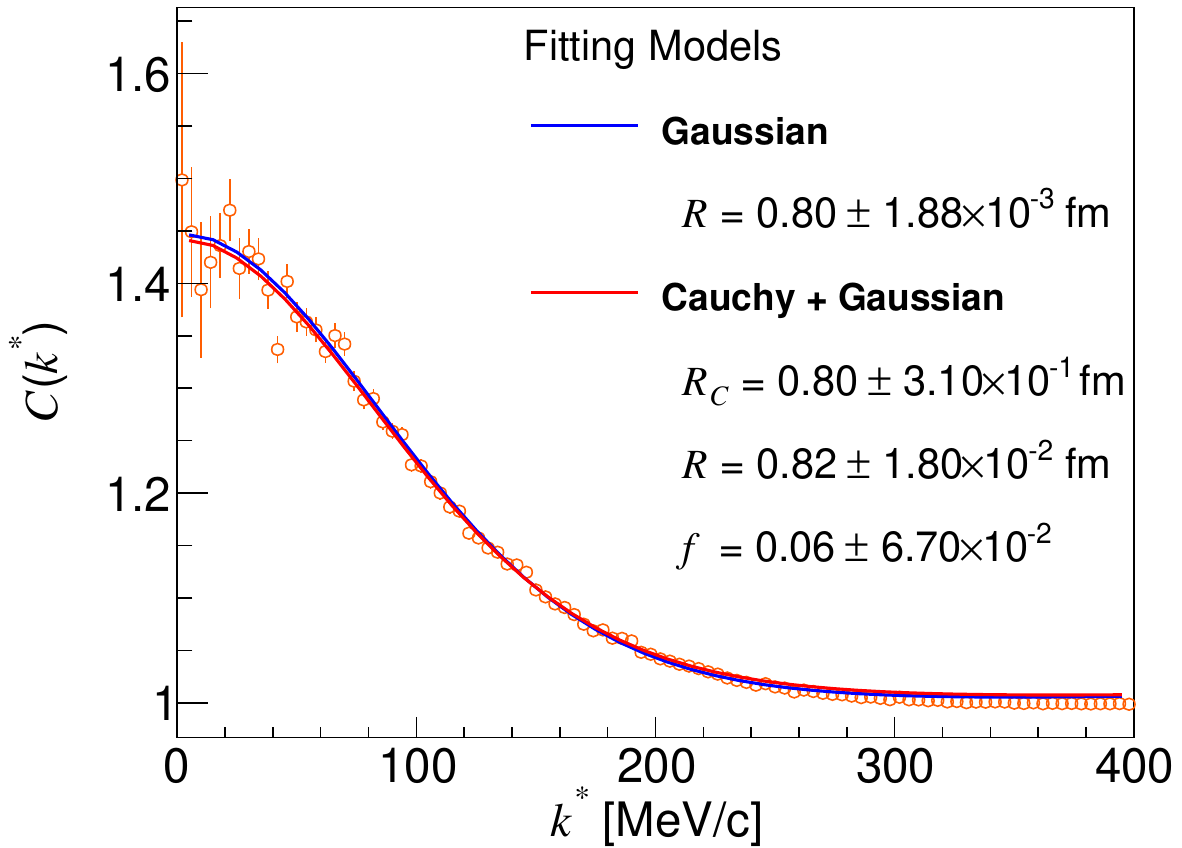}
    \caption{The $p$-$J/\psi$ correlation function (orange markers) in $\sqrt{s}=13.6~\mathrm{TeV}~pp$ collisions, {calculated} from EPOS4HQ-{simulated} particle sources and HAL QCD $N$-$J/\psi$ potentials. Results from two candidate source model fits are also shown with a same fitting range in {the} $p$-$p$ correlation. The extracted mixing fraction $f$ is found to be very small, demonstrating that the pair source is well described by a single Gaussian model.}
    \label{fig:EPOS-pJpsi-correlation}
\end{figure}

The $p$-$J/\psi$ correlation function and the corresponding source fits are presented in {Fig.} \ref{fig:EPOS-pJpsi-correlation}. Crucially, the extracted $p$-$J/\psi$ source exhibits {spatial extent
significantly smaller than that of the source of proton pairs} obtained in {Sec.} \ref{subsec:proton_source}, implying that $J/\psi$ mesons are emitted from a more compact region compared to protons. This finding provides direct motivation for applying our statistical reconstruction framework
to reconstruct the $J/\psi$ emission source from EPOS4HQ simulations. A conventional fitting strategy is used as a benchmark to compare our {methodology} ({Fig.}~\ref{fig:EPOS-pJpsi-correlation}). Since the resulting mixing fraction $f$ of {the} Gaussian + Cauchy distribution is found to be small in {describing} $p$-$J/\psi$ pair source, the pair source {can} be well described by a single Gaussian component. Adopting the {assumption} of {an} uncorrelated emission process, we extract a single-$J/\psi$ Gaussian radius of $R_{J/\psi}\simeq 0.37~\mathrm{fm}$, with a single-proton Gaussian source fitted in {Fig.}~\ref{fig:fit-proton-source}.

As discussed in {Sec.} \ref{subsec:method}, a key distinction between experiment and simulation lies in the nature of the $\tilde {C}_{k^*;i}$ observable. Experimentally measured values {[Eq.~\eqref{eq:measure_Ctilde}]} are discrete, whereas those computed from simulations {[Eq.~\eqref{eq:single-particle-correlation}]} are continuous. To emulate the discrete character of experimental data, we discretize the simulated continuous $\tilde{C}_{k^*}$ value via
\begin{equation}
\label{eq:discretization}
    n_i^\text{same}=\left\lfloor \frac{\tilde {C}_{k^*} N^\text{mixed}(k^*)}{\mathcal{N}N^\text{rare}_\text{events}(k^*)}\right\rfloor,
\end{equation}
where, $\lfloor x\rfloor$ denotes the floor function, rounding $x$ down to the nearest integer. Within the test-particle method used here, the normalization factor $\mathcal{N}=1$. The number of uncorrelated pairs $N^\text{mixed}(k^*)$ corresponds to the total count of $p$-$J/\psi$ pairs at {the} momentum $k^*$ in the EPOS4HQ-generated final-state phase space. $N^\text{rare}_\text{events}$ denotes the number of events containing one $J/\psi$. Because the $J/\psi$ yield is rare, the number of $J/\psi$ particles {[Eq.~\eqref{eq:measure_Ctilde}]} is approximately equal to the number of events containing $J/\psi$ in {Eq.}~\eqref{eq:discretization}. Therefore, fluctuations associated with eventwise particle multiplicity are strongly suppressed. Substituting $n_i^\text{same}$ back into {Eq.} \eqref{eq:measure_Ctilde} yields the discretized $\tilde{C}_{k^*;i}$, which then serves as the input for reconstructing the $J/\psi$ emission source using the {statistical reconstruction method}.

\begin{figure}
    \centering
    \includegraphics[width=0.9\linewidth]{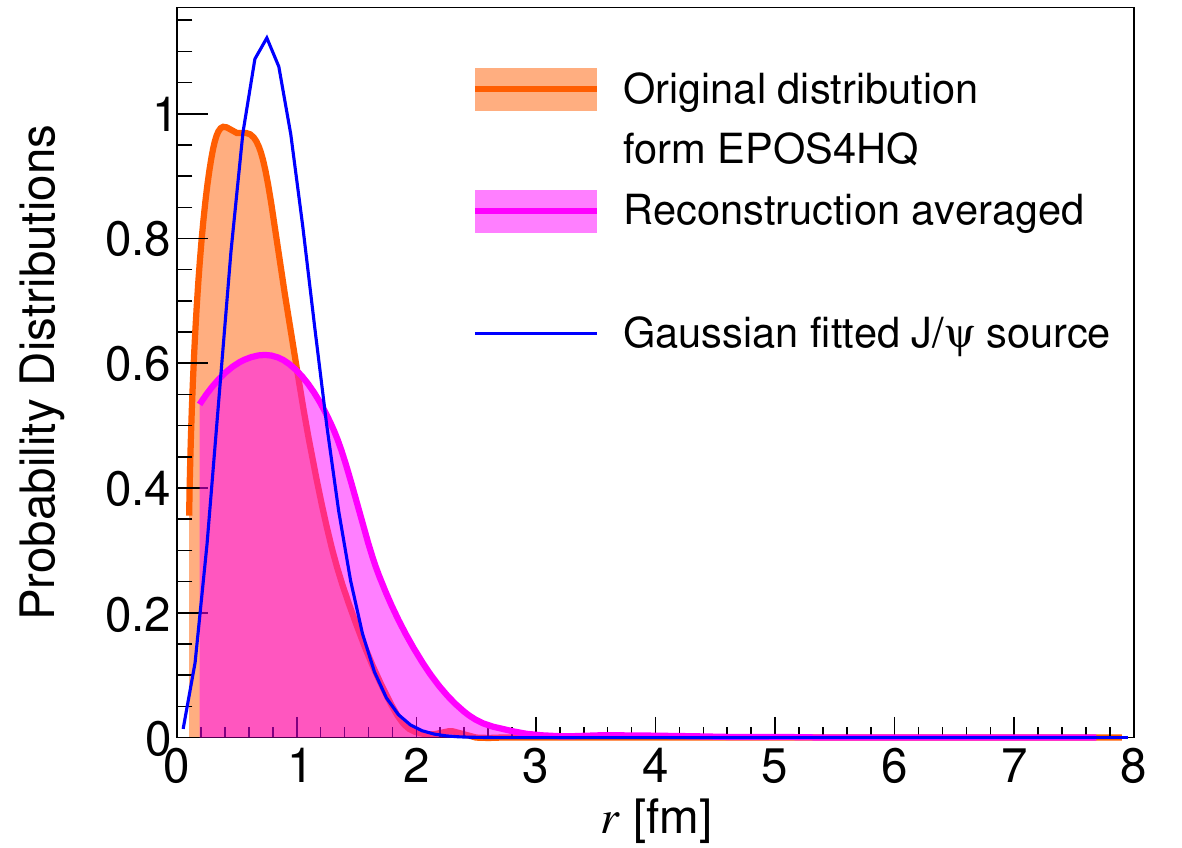}
    \caption{Comparison between {the} original $J/\psi$ source and {the} reconstruction result averaged over $k^* = {82-238~\mathrm{MeV}/c}$. The Gaussian parametrizations of {the} $J/\psi$ source extracted from {Fig.} \ref{fig:EPOS-pJpsi-correlation} are also displayed  as a comparison.}
    \label{fig:reconstruction-Jpsi}
\end{figure}

Following the methodology  in {Sec.} \ref{sec:method}, we analytically compute $\tilde{C}_{k^*}(r)$ over $r\in[0,15]$~fm. The resulting two-dimensional domain of the function on the $r$-$y$ plane is then discretized into 40 bins along the $r$ axis and coarsely binned along the $y$ axis. The coarse $y$-binning is chosen to accommodate the spread of the discrete $\tilde{C}_{k^*;i}$ values and is determined by the range of the analytically computed $\tilde{C}_{k^*}(r)$. 

{Figure} \ref{fig:reconstruction-Jpsi} shows the reconstructed $J/\psi$ emission source averaged over the momentum range $k^*={82,\dots,238~\mathrm{MeV}/c}$. While the reconstruction exhibits a broader spatial distribution and a more pronounced tail compared to the original EPOS4HQ-simulated $J/\psi$ source, it preserves key characteristics such as the most probable radius and the overall compactness. Notably, the most probable radius agrees better with the original simulation than that obtained from a simple Gaussian fit to the $J/\psi$ source. 

\begin{figure}
    \centering
    \includegraphics[width=1\linewidth]{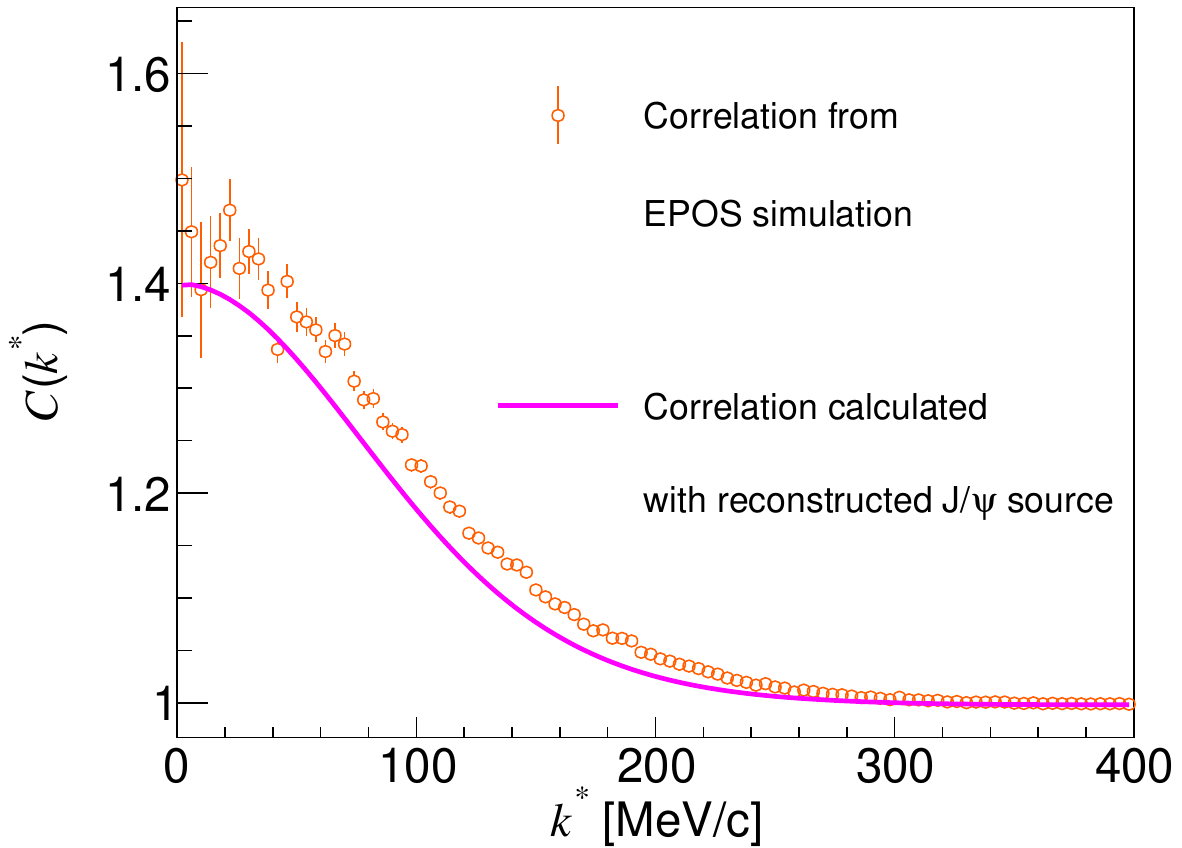}
    \caption{Comparison of the $p$-$J/\psi$ correlation function obtained from direct EPOS4HQ simulation with the result reconstructed using the {statistical reconstruction method}. 
    The momentum interval used in the multi-momentum averaging ranges from 82 to 168 ${\mathrm{MeV}/c}$ for $k^*$. This comparison is used to estimate the systematic uncertainty of the method, yielding an uncertainty of approximately $13\%$ in {the} present analysis.}
    \label{fig:reconstruction-correlation}
\end{figure}

In line with our initial conjecture in {Sec.}~\ref{sec:intro}, the reconstructed $J/\psi$ source is more compact than the proton source. Using the fitted proton source together with the reconstructed $J/\psi$ source, we compute the corresponding correlation function (magenta curve in {Fig.} \ref{fig:reconstruction-correlation}). This reconstructed correlation is weaker than the one derived directly from the EPOS4HQ {simulation}. The discrepancy originates from the broader spread and enhanced tail structure present in the reconstructed source ({Fig.} \ref{fig:reconstruction-Jpsi}). 

The systematic uncertainty of the 
{statistical reconstruction method}
is estimated via
\begin{equation}
    \label{eq:sysm_error}
    \delta_\text{sys.}\sim\frac{C_\text{recon.}(k^*)-1}{C(k^*)-1},
\end{equation}
where $C(k^*)$ denotes the $p$-$J/\psi$ correlation function computed directly from EPOS4HQ simulation, and $C_\text{recon.}(k^*)$ is the result obtained from our reconstruction. This yields a systematic uncertainty of approximately $13\%$ for this $J/\psi$ source reconstruction. One contributor to this systematic error is the discretization of the $\tilde{C}_{k^*}$ values, as well as the simple Gaussian parametrization used for the proton source. 

\section{Summary}
\label{sec:summary}
The present study presents a 
{statistical reconstruction method} 
for reconstructing the emission source of rare particles without the Gaussian parametrization.
The method is based on the particle-particle distribution of the single-particle-conditioned correlation kernel, 
which is {more easily accessible in experiments} when at most one rare particle is produced per collision event and {the fluctuations associated with particle multiplicity are suppressed}. This key property enables a direct statistical reconstruction of the source of single particles, rather than an indirect inference from effective pair sources. 

{By leveraging} a well-characterized source of reference particles and a constrained interaction potential, the conditioned kernel function 
$\tilde{C}_{k^*}(r)$ is computed analytically. Its radial derivative defines {the} momentum-dependent reconstruction kernel 
$K_{k^*}$, through which the spatial extent of the target particle source is obtained from the measured kernel distribution. 
To suppress reconstruction artifacts arising from nonmonotonic $\tilde{C}_{k^*}(r)$ behavior as {$r$ increases}, a multimomentum averaging procedure is employed as a stabilization strategy.

We conduct two {validation} studies, Gaussian source reconstruction and EPOS4HQ-simulated $J/\psi$ source reconstruction, to evaluate the method and quantify its systematic uncertainties. In the Gaussian case, Monte Carlo sampling is used to generate the spatial distribution and compute the $\tilde{C}_{k^*}$ distribution, with multi-momentum averaging yielding an accurate reconstruction. For the more complex EPOS4HQ validation, we use $p$-$J/\psi$ correlations to reconstruct {the} $J/\psi$ source. The proton source, extracted by fitting spatial distribution of protons with momentum $<400{~\mathrm{MeV}/c}$ simulated in EPOS4HQ, serves as the known reference source input. Meanwhile, the HAL QCD-derived $N$-$J/\psi$ potentials are used to {calculate} the $p$-$J/\psi$ correlation and the conditioned kernel 
$\tilde{C}_{k^*}$. {By applying} multimomentum averaging, {the key characteristics of the reconstructed $J/\psi$ source are preserved}, including the most probable radius and intrinsic compactness. A comparison between the directly simulated and reconstructed $p$-$J/\psi$ correlation functions reveals  a systematic uncertainty of approximately $13\%$ for $J/\psi$ source reconstruction.

In summary, the proposed {statistical reconstruction method} 
enables reconstruction of rare-particle emission sources without  assumptions on the shape. This approach could provide quantitative constraints on source characteristics of rare particles and offer deeper insights into the intermediate processes of both $pp$ and heavy-ion collisions. The methodology can further serve as a useful tool for future femtoscopy analysis.

\begin{acknowledgments}
This work was supported in part by the National Natural Science Foundation of China under Contract No. 12275054, No. 12147101, No. 12061141008, No. 12347106, No. 12422509,  No. 12375121, and No. 12547102, the National Key R\&D Program of China under Granta No. 2024YFA1610802 and No. 2018YFE0104600, the Guangdong Major Project of Basic and Applied Basic Research under Granta No. 2020B0301030008, the Shanghai Pilot Program for Basic Research--Fudan University under Granta No.  21TQ1400100(22TQ006) and the STCSM under Grant No. 23590780100.
\end{acknowledgments}
\end{CJK*} 
\bibliographystyle{apsrev4-2}
\bibliography{03}

\end{document}